\documentclass{pasj00}
\draft

\begin{document}
\SetRunningHead{Momose et al.}{Disk around HD169142 in \textit{H}-band}

\title{Detailed structure of the outer disk 
around HD 169142 with polarized light in $H$-band}

%

%
 \author{%
   Munetake \textsc{Momose}\altaffilmark{1}\thanks{E-mail:~momose@mx.ibaraki.ac.jp},
   Ayaka \textsc{Morita}\altaffilmark{1},
   Misato \textsc{Fukagawa}\altaffilmark{2},
   Takayuki \textsc{Muto}\altaffilmark{3},
   Taku \textsc{Takeuchi}\altaffilmark{4},
   Jun \textsc{Hashimoto}\altaffilmark{5},
   Mitsuhiko \textsc{Honda}\altaffilmark{6},
   Tomoyuki \textsc{Kudo}\altaffilmark{7},
   Yoshiko K. \textsc{Okamoto}\altaffilmark{1},
   Kazuhiro D. \textsc{Kanagawa}\altaffilmark{8},
   Hidekazu \textsc{Tanaka}\altaffilmark{8},
   Carol A. \textsc{Grady}\altaffilmark{9,10,11},
   Michael L. \textsc{Sitko}\altaffilmark{12,13}\thanks{Visiting Astronomer at the Infrared Telescope Facility, which is operated by the University of Hawaii under Cooperative Agreement no. NNX-08AE38A with the National Aeronautics and Space Administration, Science Mission Directorate, Planetary Astronomy Program.}, 
   Eiji \textsc{Akiyama}\altaffilmark{14}, 
   Thayne \textsc{Currie}\altaffilmark{7},
   Katherine B. \textsc{Follette}\altaffilmark{15}, 
   Satoshi \textsc{Mayama}\altaffilmark{16,17}, 
   Nobuhiko \textsc{Kusakabe}\altaffilmark{14}, 
   Lyu \textsc{Abe}\altaffilmark{18}, 
   Wolfgang \textsc{Brandner}\altaffilmark{19}, 
   Timothy D. \textsc{Brandt}\altaffilmark{20}, 
   Joseph C. \textsc{Carson}\altaffilmark{21}, 
   Sebastian \textsc{Egner}\altaffilmark{7}, 
   Markus \textsc{Feldt}\altaffilmark{19}, 
   Miwa \textsc{Goto}\altaffilmark{22}, 
   Olivier \textsc{Guyon}\altaffilmark{7}, 
   Yutaka \textsc{Hayano}\altaffilmark{7}, 
   Masahiko \textsc{Hayashi}\altaffilmark{14}, 
   Saeko S. \textsc{Hayashi}\altaffilmark{7}, 
   Thomas \textsc{Henning}\altaffilmark{19}, 
   Klaus W. \textsc{Hodapp}\altaffilmark{23}, 
   Miki \textsc{Ishii}\altaffilmark{14}, 
   Masanori \textsc{Iye}\altaffilmark{14}, 
   Markus \textsc{Janson}\altaffilmark{24}, 
   Ryo \textsc{Kandori}\altaffilmark{14}, 
   Gillian R. \textsc{Knapp}\altaffilmark{25}, 
   Masayuki \textsc{Kuzuhara}\altaffilmark{4}, 
   Jungmi \textsc{Kwon}\altaffilmark{26}, 
   Taro \textsc{Matsuo}\altaffilmark{27}, 
   Michael W. \textsc{McElwain}\altaffilmark{9}, 
   Shoken \textsc{Miyama}\altaffilmark{28}, 
   Jun-Ichi \textsc{Morino}\altaffilmark{14}, 
   Amaya \textsc{Moro-Martin}\altaffilmark{25,29}, 
   Tetsuo \textsc{Nishimura}\altaffilmark{7}, 
   Tae-Soo \textsc{Pyo}\altaffilmark{7}, 
   Eugene \textsc{Serabyn}\altaffilmark{30}, 
   Takuya \textsc{Suenaga}\altaffilmark{14,17}, 
   Hiroshi \textsc{Suto}\altaffilmark{14}, 
   Ryuji  \textsc{Suzuki}\altaffilmark{14}, 
   Yasuhiro H. \textsc{Takahashi}\altaffilmark{14,26}, 
   Michihiro \textsc{Takami}\altaffilmark{31}, 
   Naruhisa \textsc{Takato}\altaffilmark{7},
   Hiroshi \textsc{Terada}\altaffilmark{7}, 
   Christian \textsc{Thalmann}\altaffilmark{32}, 
   Daigo \textsc{Tomono}\altaffilmark{7}, 
   Edwin L. \textsc{Turner}\altaffilmark{25,33}, 
   Makoto \textsc{Watanabe}\altaffilmark{34}, 
   John \textsc{Wisniewski}\altaffilmark{5}, 
   Toru \textsc{Yamada}\altaffilmark{35}, 
   Hideki \textsc{Takami}\altaffilmark{14},
   Tomonori \textsc{Usuda}\altaffilmark{14}, 
   and
   Motohide \textsc{Tamura}\altaffilmark{14,26}}
   
 \altaffiltext{1}{College of Science, Ibaraki University, 2-1-1 Bunkyo, Mito, Ibaraki 310-8512}
 \altaffiltext{2}{Graduate School of Science, Osaka University, 1-1 Machikaneyama, Toyonaka, Osaka 560-0043}
 \altaffiltext{3}{Division of Liberal Arts, Kogakuin University, 1-24-2 Nishi-Shinjuku, Shinjuku-ku, Tokyo 163-8677}
 \altaffiltext{4}{Department of Earth and Planetary Sciences, Tokyo Institute of Technology, Meguro-ku, Tokyo 152-8551}
 \altaffiltext{5}{H. L. Dodge Department of Physics and Astronomy, University of Oklahoma, 440 W Brooks St Norman, OK 73019, USA}
 \altaffiltext{6}{Department of Mathematics and Physics, Kanagawa University, 2946 Tsuchiya, Hiratsuka, Kanagawa 259-1293}
 \altaffiltext{7}{Subaru Telescope, 650 North A'ohoku Place, Hilo, HI 96720, USA}
 \altaffiltext{8}{Institute of Low Temperature Science, Hokkaido University, Sapporo, Hokkaido 060-0819}
 \altaffiltext{9}{Exoplanets and Stellar Astrophysics Laboratory, Code 667, Goddard Space Flight Center, Greenbelt, MD 20771, USA}
 \altaffiltext{10}{Eureka Scientific, 2452 Delmer, Suite 100, Oakland, CA 96002, USA}
 \altaffiltext{11}{Goddard Center for Astrobiology}
 \altaffiltext{12}{Space Science Institute, 4750 Walnut St., Suite 205, Boulder, CO 80301, USA}
 \altaffiltext{13}{Department of Physics, University of Cincinnati, Cincinnati, OH 45221-0011, USA}
 \altaffiltext{14}{National Astronomical Observatory of Japan, 2-21-1 Osawa, Mitaka, Tokyo 181-8588}
 \altaffiltext{15}{Steward Observatory, University of Arizona, 933 N Cherry Ave, Tucson, AZ 85721, USA}
 \altaffiltext{16}{The Center for the Promotion of Integrated Sciences, The Graduate University for Advanced Studies~(SOKENDAI), Shonan International Village, Hayama-cho, Miura-gun, Kanagawa 240-0193}
 \altaffiltext{17}{Department of Astronomical Science, The Graduate University for Advanced Studies ~ (SOKENDAI), 2-21-1 Osawa, Mitaka, Tokyo 181-8588}
 \altaffiltext{18}{Laboratoire Lagrange (UMR 7293), Universit\`e de Nice-Sophia Antipolis, CNRS, Observatoire de la C\^ote d'Azur, 28 avenue Valrose, 06108 Nice Cedex 2, France}
 \altaffiltext{19}{Max Planck Institute for Astronomy, K\"onigstuhl 17, 69117 Heidelberg, Germany}
 \altaffiltext{20}{Astrophysics Department, Institute for Advanced Study, Princeton, NJ 08540, USA}
 \altaffiltext{21}{Department of Physics and Astronomy, College of Charleston, 66 George St., Charleston, SC 29424, USA}
 \altaffiltext{22}{Universit\"ats-Sternwarte M\"unchen, Ludwig-Maximilians-Universit\"at, Scheinerstr. 1, 81679 M\"unchen, Germany}
 \altaffiltext{23}{Institute for Astronomy, University of Hawaii, 640 N. A'ohoku Place, Hilo, HI 96720, USA}
 \altaffiltext{24}{Department of Astronomy, Stockholm University, AlbaNova University Center,
106 91 Stockholm, Sweden}
 \altaffiltext{25}{Department of Astrophysical Science, Princeton University, Peyton Hall, Ivy Lane, Princeton, NJ 08544, USA}
 \altaffiltext{26}{Department of Astronomy, The University of Tokyo, 7-3-1 Hongo, Bunkyo-ku, Tokyo 113-0033}
 \altaffiltext{27}{Department of Astronomy, Kyoto University, Kitashirakawa-Oiwake-cho, Sakyo-ku, Kyoto, Kyoto 606-8502}
 \altaffiltext{28}{Hiroshima University, 1-3-2, Kagamiyama, Higashihiroshima, Hiroshima 739-8511}
 \altaffiltext{29}{Department of Astrophysics, CAB-CSIC/INTA, 28850 Torrej\'on de Ardoz, Madrid, Spain}
 \altaffiltext{30}{Jet Propulsion Laboratory, California Institute of Technology, 4800 Oak Grove Drive, Pasadena, CA 91109, USA}
 \altaffiltext{31}{Institute of Astronomy and Astrophysics, Academia Sinica, P.O. Box 23-141, Taipei 10617, Taiwan}
 \altaffiltext{32}{Astronomical Institute ``Anton Pannekoek'', University of Amsterdam, Postbus 94249, 1090 GE, Amsterdam, The Netherlands}
 \altaffiltext{33}{Kavli Institute for Physics and Mathematics of the Universe, The University of Tokyo, 5-1-5, Kashiwanoha, Kashiwa, Chiba 277-8568}
 \altaffiltext{34}{Department of Cosmosciences, Hokkaido University, Kita-ku, Sapporo, Hokkaido 060-0810}
 \altaffiltext{35}{Astronomical Institute, Tohoku University, Aoba-ku, Sendai, Miyagi 980-8578}

\maketitle

\begin{abstract}
Coronagraphic imagery of the circumstellar disk 
around HD 169142 in $H$-band polarized intensity (PI) 
with Subaru/HiCIAO is presented.  The emission scattered by dust 
particles at the disk surface in $0\farcs2 \leq r \leq 1\farcs2$,
or $29 \leq r \leq 174$ AU, is successfully detected.  
The azimuthally-averaged radial profile of the PI shows 
a double power-law distribution, in which the PIs in 
$r=29-52$ AU and $r=81.2-145$ AU respectively show 
$r^{-3}$-dependence. These two power-law regions are 
connected smoothly with a transition zone (TZ), 
exhibiting an apparent gap in $r=40-70$ AU.
The PI in the inner power-law region shows a deep 
minimum whose location seems to coincide with the point 
source at $\lambda = 7$ mm. 
This can be regarded as another sign of a protoplanet in TZ.  
The observed radial profile of the PI is reproduced 
by a minimally flaring disk with
an irregular surface density distribution 
or with an irregular temperature distribution 
or with the combination of both.
The depletion factor of 
surface density in the inner power-law region ($r< 50$ AU) 
is derived to be $\geq 0.16$ from a simple model calculation. 
The obtained PI image also shows small scale asymmetries 
in the outer power-law region. 
Possible origins for these asymmetries include corrugation of the
scattering surface in the outer region, and shadowing effect by a puffed
up structure in the inner power-law region.
\end{abstract}

\section{Introduction}

Circumstellar disks around young stars are the key targets 
to obtain a better understanding of the formation of planetary systems 
\citep{araa10,araa11}.
During ongoing formation of a planet, the disk is expected to manifest 
observable signatures inside, such as an inner hole, a gap, spirals and 
other asymmetric patterns. 
Recent progress in the technique of differential imaging 
at near infrared wavelengths finally allows 
us to detect small-scale complex features in the disks (e.g., 
\cite{hashi11,hashi12,muto12,mayama12,grady13,follette13}). 
Most of these objects are classified as transitional disks, which 
show significant amount of excess emission at wavelengths $\lambda \gtrsim 10~\mu$m 
but little excess at shorter wavelengths, suggesting that dust  in the
inner regions has already been depleted \citep{strom89,skur90}. 
Imaging surveys of transitional disks in dust thermal emission
at submillimeter wavelengths clearly show that these disks do commonly 
possess inner holes \citep{brown09,andrews11}. Recently, even more complex
distributions of dust emission on a smaller size scale are also revealed
in some of these objects (e.g., \cite{cass13,vdm13,fukagawa13,perez14}). 
These  asymmetric features may be caused by trapping of dust particles in a vortex 
with a higher pressure, which can lead to efficient formation of rocky planetesimals
\citep{regaly12,birnstiel13}. 

HD 169142 is a Herbig Ae star located at 145 pc from the Sun \citep{sylv96},  
and the spectral type of the star is classified as A5Ve (Dunkin, Barlow \& Ryan 1997) 
or A8Ve \citep{grady07}. The age is estimated to be $6^{+6}_{-3}$ Myr 
from the ages of two pre-main sequence stars 
located $9\farcs3$ to the southwest and are comoving with HD 169142 \citep{grady07}. 
The luminosity of $9.4 L_{\odot}$ is used by \citet{meeus12}, 
while \citet{manoj06} derive $18.2 L_{\odot}$ with $A_V = 0.61$ 
(see also \cite{osorio14}). 
Judging from the observed $B-V$ color 0.26, however, $A_V\approx 0.5$ 
was the case only if the stellar spectral type was as early as B8. 
\citet{wagner14}, in fact, find a 
good fit to the SED, including the UV data, for 
the spectral type of A7, 
the effective temperature of 7500K and the stellar radius of 
$1.7 R_{\odot}$ (resulting in $8.7L_{\odot}$) with no foreground extinction.
It exhibits significant amount of excess emission at longer wavelengths and 
its spectral energy distribution (SED) is categorized as group I \citep{meeus01}. 
Interferometric observations of CO lines at millimeter 
wavelengths revealed a gas disk with the radius of $2\arcsec~(\approx 300\mathrm{AU})$, 
and their velocity fields are 
consistent with Keplerian rotation 
of the central stellar mass $2 M_{\odot}$ and its inclination angle 
$13^{\circ}$ \citep{raman06,panic08}. Scattered light at near infrared wavelengths 
from the disk outer regions was also imaged with coronagraph 
instruments \citep{grady07,fukagawa10}. 

More recently, the inner regions of the disk have been investigated by 
both the analysis on its broadband SED and higher resolution imagery.  
It was claimed that there should be an inner cavity around HD 169142
to reproduce its SED, but its size was rather unclear 
solely from the SED analysis \citep{grady07,meeus10}. 
\citet{honda12} carried out imaging of the disk at $\lambda =18.8$ and 
24.5~$\mu$m with Subaru/COMICS, and by combining these data 
with a two-dimensional radiative transfer model for the 
SED, they determined the location of the inner wall of the disk to be 
$23\pm3$ AU from the star. They also speculate that the SED
classification as group I may correspond to the object accompanied by a
(pre)transitional disk in which there is a large inner hole 
and the inner wall of the disk just outside 
the hole can account for its characteristic large amount of excess 
emission at mid-infrared wavelengths (see also \cite{maas13}). 
\citet{quanz13} made observations 
of polarization intensity (PI) in $H$-band ($\lambda = 1.65~\mu$m) with VLT/NACO . 
They found a bright ring of $r\approx 25$ AU that agreed 
with the estimate from the SED analyses \citep{meeus10,honda12}, 
as well as an annular gap extending $r=40-70$ AU. 
Furthermore, imaging of dust thermal radiation at $\lambda = 7$ mm 
with J-VLA also shows asymmetric ring-like emission 
at $r\approx 25$ AU and a hint of an outer gap at $40-70$ AU, suggesting that 
the structure seen in the scattered light at near infrared wavelengths 
may correspond to real variations in surface density \citep{osorio14}. 
Although these are not the decisive evidence for the existence of 
protoplanet(s), such a gap is reminiscent of that 
induced by a planet in the disk 
(e.g., \cite{lin93}; Crida, Morbidelli \& Masset 2006). 
The latest observations in $L^\prime$-band ($\lambda = 3.8~\mu$m) with 
an annular groove phase mask (AGPM) vector-vortex coronagraph mounted on VLT 
detect a companion candidate inside the inner cavity ($0\farcs156$, 
or 23 AU from the star), which might be the direct detection of 
radiation from a protoplanet \citep{biller14,reggiani14}. 

This paper presents coronagraphic imagery of the circumstellar disk 
around HD 169142 in $H$-band polarized intensity (PI) 
with Subaru/HiCIAO. 
We describe the details of the observations in \S 2
and present the results in \S 3. 
Based on our results as well as those obtained by previous studies, 
we discuss in \S 4 the disk structure and estimate the nature of 
a possible protoplanet embedded in the disk from our results. 

\section{Observations and Data Reduction}

\subsection{Observations with Subaru/HiCIAO}

HD 169142 was observed in $H$-band ($\lambda = 1.65~\mu$m) with 
the high-contrast imaging instrument HiCIAO 
\citep{tamura06,hodapp08,suzuki10}
on the Subaru Telescope on 2011 May 23 UT as part of 
Strategic Explorations of Exoplanets and Disks with Subaru 
(SEEDS; \cite{tamura09}).
The data were taken in a combined angular differential imaging (ADI) 
and polarization differential imaging (PDI) mode with a field of 
view of $10''\times 20''$ and a pixel scale of 9.5 milli-arcsecond (mas).
A circular occulting mask $0\farcs15$ in radius was used to suppress 
the bright stellar halo. The half-wave plates were periodically placed 
at four angular positions from $0^{\circ}, 45^{\circ}, 22.5^{\circ}$ 
and $67.5^{\circ}$ in sequence with 30s exposure per wave 
plate position. The total integration time of the PI image was 
1080s after removing low quality frames with large FWHMs
by careful inspection of the stellar point-spread function (PSF).
Data of the reference star HD 166903, whose apparent magnitude in 
$H$-band is 5.837 mag (SIMBAD), were also taken in the same observing mode. 
It was proved that a stable stellar PSF of 
FWHM = $0\farcs06$ was achieved 
for the selected images of HD 169142 by the adaptive optics system 
AO188 \citep{hayano04,minowa10}.

\subsection{Data reduction}

The data reduction procedure was the same as that 
described by \citet{muto12}. 
The raw images were first corrected with 
IRAF\footnote{IRAF is 
distributed by the National Optical Astronomy Observatory, which 
is operated by the Association of Universities for Research in Astronomy, Inc., 
under cooperative agreement with the National Science Foundation.} 
for bias and dark current, and flat-fielding was 
performed after sky subtraction.
We applied a distortion correction using globular cluster M5 data 
taken within a few days, by IRAF packages GEOMAP and GEOTRAN. 
The images of Stokes ($Q, U$) parameters 
were obtained in the standard 
way (e.g., \cite{hinkley09}); 
By subtracting two images of extraordinary and ordinary rays 
at each wave plate position, we got $+Q, -Q, +U$, and $-U$ 
images, from which $2Q$ and $2U$ images were made by another 
subtraction to eliminate remaining aberration. 
Instrumental polarization of 
HiCIAO was corrected  by the way described by 
\citet{joos08}. 

We needed to fix the rotation center of the field of view by ADI 
in de-rotating the $Q$ and $U$ images. It was determined in the following way. 
A pair of $I$-frames that were taken more than 300s apart from each other 
was picked up, and then the rotation center and rotation angle for the pair were 
derived so that the two background stars in the differential image between 
the two frames were best cancelled out. These values were collected for all 
the possible $I$-frame pairs, and their average position 
was regarded as the image rotation center in our observations. 
There were four background stars in each frame, but two of them were not used 
because their positions were so far from the image center that the distortion correction 
might be imperfect. It proved that the mask center position was shifted from 
the image rotation center by ($+1, -4$) pixels in ($\alpha, \delta$), or 
($+9.5$ mas, $-38$ mas). 
On the other hand, 
the stellar position was estimated from elliptical fitting to an iso-intensity 
contour of the periphery of stellar halo in $I$-frames. 
The stellar location measured in this way was proved to be the same as 
the image rotation center. The $Q$ and $U$ images were de-rotated at this center, 
then averaged to increase the signal-to-noise. PI was then obtained by 
PI = $\sqrt{Q^2+U^2}$. The misalignment between the mask center 
and image rotation center made the resultant working angle 
$\geq 0\farcs2$ from the star, which is slightly larger than the actual mask size.

\section{Results \label{results}}

\begin{figure}
  \begin{center}
    \includegraphics[width=14cm]{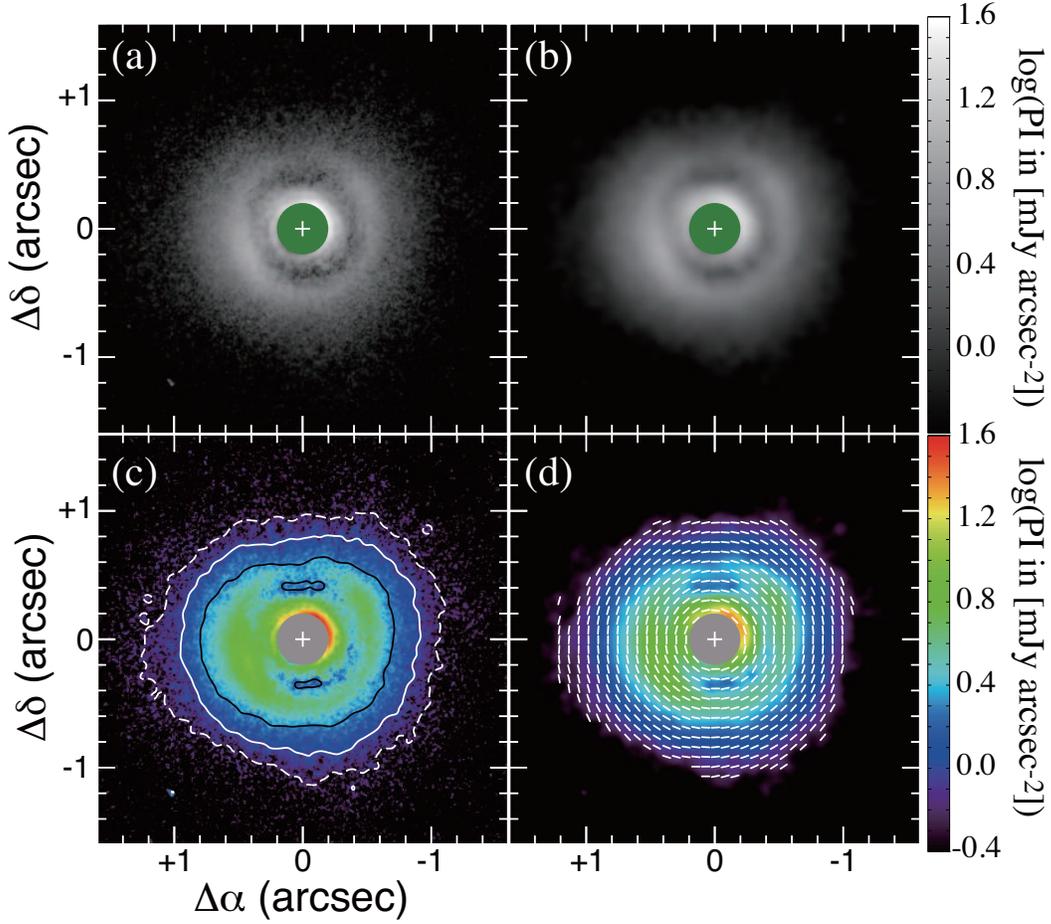} 
   \end{center}
\caption{(a) Coronagraphic image of the polarization intensity (PI) 
in $H$-band around HD 169142 obtained with Subaru/HiCIAO.
White cross is the stellar position, and the green circle shows the regions affected by 
the mask, or $< 0.2\arcsec$ from the stellar position. 
(b) PI image convolved with a Gaussian function whose FWHM is 8 pixels, 
or 0\farcs072. 
(c) Same as (a), but in different color code so that asymmetric features 
are easily identified. White broken lines are the contours of 0.5 mJy arcsec$^{-2}$, 
corresponding 
to the 3$\sigma$ level (see the text). Solid lines in white and black are the 
contours at 1.0 mJy arcsec$^{-2}~(6\sigma)$ and 1.5 mJy arcsec$^{-2}~(9\sigma)$ levels, 
respectively. The contours are generated in the convolved image shown in (b). 
(d) Same as (b), but in the color code as (c), superposed on 
the polarization vectors where the PI is detected above the 3$\sigma$ level. 
The polarization vectors are drawn every 8 pixels.}\label{polimage}
\end{figure}

Figure \ref{polimage} shows the PI image of HD 169142 in $H$-band with 
Subaru/HiCIAO. 
The infrared emission scattered by dust particles at the disk surface is
successfully detected. 
The typical PI value is $\approx 0.21$ mJy arcsec$^{-2}$ in $r > 3\arcsec$, 
well beyond the disk outer radius derived from the CO observations 
\citep{raman06,panic08} and the scattered light of the disk should be negligible.
This corresponds to the $1\sigma$ level of 0.168 mJy arcsec$^{-2}$ if 
the values in the region of no polarized emission follow 
the Rayleigh distribution.
In the following quantitative analyses, 
we will regard the PI emission of $>0.5$ mJy arcsec$^{-2}$
($= 3\sigma$ level) as a significant detection. 
The region of significant detection extends up to $r \approx 1\farcs2$ 
from the star, as presented in figure \ref{polimage}c. 
Figure \ref{polimage}d shows the polarization vectors 
derived from Stokes $Q$ and $U$ images. 
The polarization vectors align azimuthally, 
as expected for the stellar light scattered at 
the disk surface. 

The overall structure of the PI is roughly consistent with that 
obtained by \citet{quanz13} with VLT/NACO .
The obtained PI in figure \ref{polimage}a shows an 
axisymmetric distribution as follows:
(i) the bright ring just outside the mask, 
(ii) a ring-like feature at $r\approx 0\farcs55$, or 80 AU from the star, 
(iii) the apparent gap between these two ring-like features, 
and (iv) the outermost part in which the PI gets gradually 
weaker at a larger radius. 
Note that the PI is significantly detected in the gap (iii) with at least $9\sigma$.
These axisymmetric features can be seen even more clearly in the 
smoothed image convolved with a Gaussian function (figure \ref{polimage}b).  
The inner bright rim at $r\approx 0\farcs2$ 
\citep{quanz13} is only marginally confirmed  
because the inner working angle of our observation is slightly larger 
than that of VLT/NACO.

\begin{figure}
  \begin{center}
    \includegraphics[width=14cm]{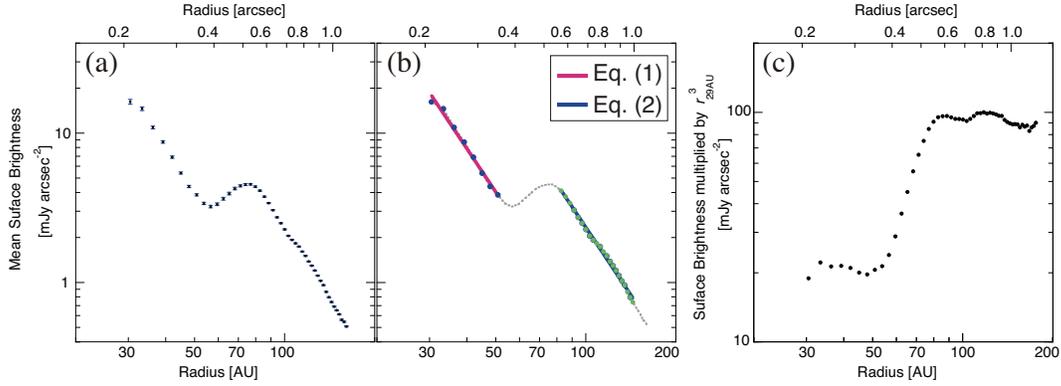}
  \end{center}
\caption{(a) Averaged radial profile of the PI above the 3$\sigma$ level (0.5 mJy arcsec$^{-2}$), 
between (29$-$159.5) AU, or $0\farcs2-1\farcs1$.
Averages and their uncertainties are calculated in annuli of 2.9AU ($=0\farcs02$) width 
after the correction for the projection on the sky under the assumptions 
that the position angle of the disk major axis is 5$^{\circ}$ ~and the disk 
inclination angle is 13$^{\circ}$. 
(b) Results of the fitting by a power-law function to the data points in 
29$-$52.2 AU (\textit{red}) and in 81.2$-$145 AU (\textit{blue}); 
the data points used in the power-law fitting are shown in blue and 
green dots.  
Radial profile in (a) is shown by gray dotted line. 
(c) Same as (a), but normalized by $r_{29\mathrm{AU}}^{-3}$, where 
 $r_{29\mathrm{AU}}$ is a dimensionless radius normalized by 29 AU.}
\label{pi-rprofile}
\end{figure}

To describe the axisymmetric distribution of the PI more quantitatively, 
its azimuthally-averaged radial profile is presented in figure \ref{pi-rprofile}; 
the projection on the sky is corrected 
under the assumptions that 
the disk inclination angle is $13^{\circ}$ and the disk major axis 
is in PA$=5^{\circ}$ \citep{raman06}. 
The radial profile shows a double power-law distribution, as expressed 
by 
\begin{equation}
\mathrm{PI}(r) =20.7\left( \frac{r}{\mathrm{29~AU}}\right)^{-3.0025}
~[\mathrm{mJy~arcsec}^{-2}] 
\label{eqn:pl1}\\
\end{equation}
in $29 \leq r \leq 52.2~\mathrm{AU}$ and 
\begin{equation}
\mathrm{PI}(r) =96.2\left( \frac{r}{\mathrm{29~AU}}\right)^{-3.0037}
~[\mathrm{mJy~arcsec}^{-2}]\label{eqn:pl2}
\end{equation}
in $81.2\leq r \leq 145~\mathrm{AU}$, respectively. 
The formal statistical uncertainties in the amplitudes 
are estimated to be $\lesssim 10$\% for (\ref{eqn:pl1}) and 
$\lesssim  1.3$\% for (\ref{eqn:pl2}), 
and those in the exponents are $\lesssim 0.1$ dex. 
These two power-law regions are connected smoothly with a
``transition zone'' (hereafter denoted by TZ) extending $r=52-81$ AU. 
The apparent gap located in 
$r\approx 40-70$ AU \citep{quanz13,osorio14} consists of the outer 
part of the inner power-law region and TZ, while the bright ring 
at $r\approx 0\farcs55$ in the PI image (figures 
\ref{polimage}a and \ref{polimage}b) 
corresponds to the inner boundary of the outer power-law region. 
It should be noted, however, that the surface brightness 
normalized by the $r^{-3}$ power-law (figure \ref{pi-rprofile}c) changes 
smoothly even in TZ.
There is no break of the exponent at $r= 120$ AU claimed by 
\citet{quanz13}, though a subtle irregularity may be identified 
at $r\approx 100$ AU.
The $r^{-3}$-dependence was also found in the total intensity
distribution at 1.1 $\mu$m in $r \geq 80$ AU with the coronagraphic 
imaging by \textit{HST}/NICMOS \citep{grady07}. 
It is also consistent with the radial exponent 
in $116 \le r \le 174$ AU derived from the imagery in $H$-band with 
Subaru/CIAO ($-3.0\pm 0.2$; \cite{fukagawa10}), suggesting that 
there is no significant radial variation of polarization degree 
in these regions.
The polarization degree estimated from the comparisons between 
the PI in figure \ref{polimage} and the total intensity by \citet{fukagawa10} 
is $\sim 65$ \% in $0\farcs8 \le r \le 1\farcs1$.
This is higher than that obtained in the $H$-band PI imaging 
of the disk around HD 142527 \citep{ave14}. It may be due to the 
nearly pole-on configuration of HD 169142; the scattering angle 
at the disk surface should be $\sim 90^{\circ}$ at which the polarization 
degree is $\sim 100$\% when the particle size is much smaller than the wavelength 
\citep{bh83}.
Alternatively, it might be due to 
an underestimate of total scattered-light intensity 
caused by a systematic error during PSF subtraction by 
\citet{fukagawa10}.
The rest of the discussion in this paper will not be affected by 
any uncertainties in polarization degree. 

\begin{figure}
  \begin{center}
    \includegraphics[width=14cm]{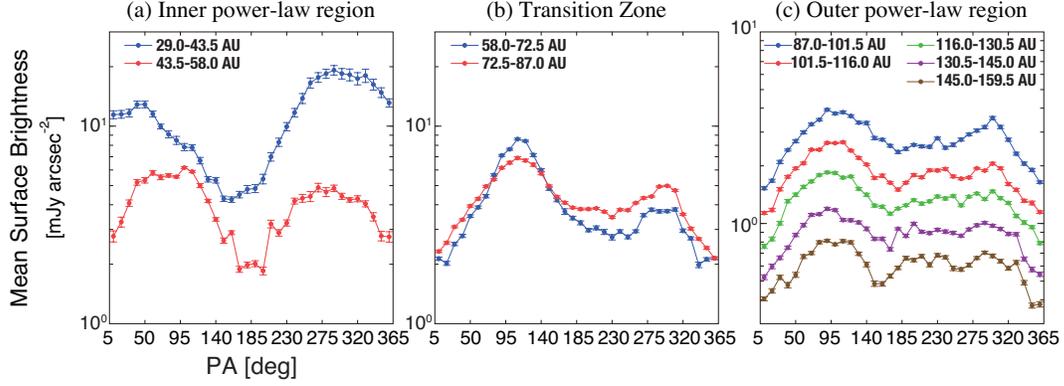}
  \end{center}
  \caption{Azimuthal distributions of the PI in annuli of
14.5AU (or 0\farcs1) width. 
The projection on the sky is corrected under the same
assumptions as those described in the caption of figure \ref{pi-rprofile}. 
The abscissa starts from PA$=5^{\circ}$, corresponding to the 
major axis of the disk. 
(a) in the inner power-law region, 
(b) in the transition zone (TZ), and
(c) in the outer power-law region.  }
\label{pi-azprofile}
\end{figure}

\begin{figure}
  \begin{center}
    \includegraphics[width=12cm]{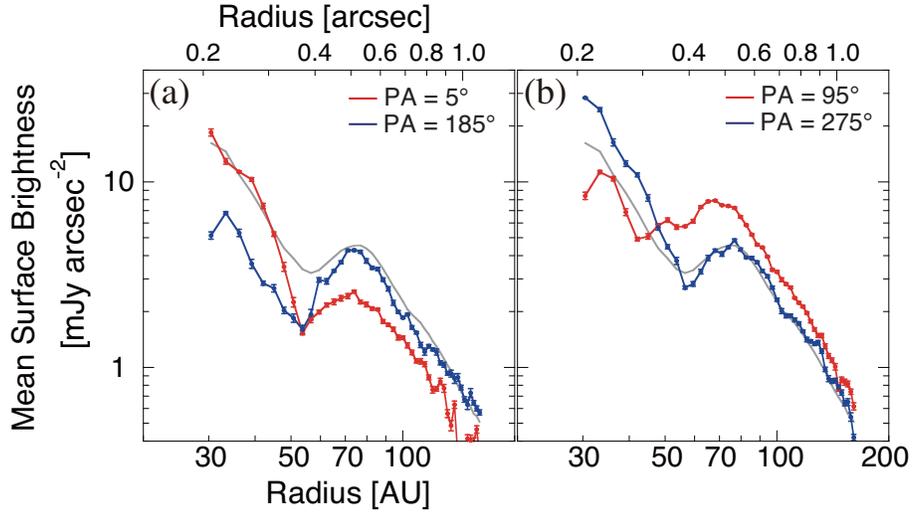}
  \end{center}
  \caption{(a) Radial profiles of the PI along the major 
  axis of the disk: PA = $5^{\circ} \pm 5^{\circ}$ in red and 
  PA = $185^{\circ} \pm 5^{\circ}$ in blue. (b) Radial profiles of the PI 
  along the minor axis of the disk: PA = $95^{\circ} \pm 5^{\circ}$ in red 
  and PA = $275^{\circ} \pm 5^{\circ}$ in blue. Radial profile averaged 
  over the whole direction (shown in figure \ref{pi-rprofile}) is indicated 
  by gray line in each panel. }
\label{pi-majmin}
\end{figure}

The PI image also manifests asymmetric features, which are easily 
identified in figure \ref{polimage}c. 
To see the asymmetry at each radius, 
azimuthal distributions of the PI in annuli of 14.5 AU width
are presented in figure \ref{pi-azprofile}.
In the innermost annulus (29.0 $\leq r \leq$ 43.5 AU), 
the northwestern part ($260^{\circ} \lesssim \mathrm{PA} \lesssim 340^{\circ}$)
is quite bright. This may be a part of the inner rim distorted outwardly 
\citep{quanz13}. 
Except for this innermost annulus, all the azimuthal distributions have their maxima at 
PA$\approx 110^{\circ}$, close to the disk minor axis east to the star. 
This may imply that the eastern part is the near side of the disk and is 
brighter than the western side because of the forward scattering. 
In the outer power-law region, 
the PI near the major axis ($\mathrm{PA} = 5^{\circ}$ and $185^{\circ}$) 
is dimmer than those near the minor axis 
($\mathrm{PA} = 95^{\circ}$ and $275^{\circ}$), 
which is also found in figure \ref{pi-majmin}. 
As a consequence, the lowest contour of the significant detection in 
figure \ref{polimage}c is slightly elongated in east-west direction.
This situation is completely different from other disks 
in scattered light detected with Subaru/HiCIAO, 
in which the PI is brighter along the major axis 
(e.g., \cite{muto12, kusakabe12}). 
The variation in each annulus is systematically larger at 
inner radii; In the inner power-law region and TZ, the amount of 
variation in each annulus is $4.5-3.3$, but those of annuli in the 
outer power-law region is less than $2.6$.

\begin{figure}
  \begin{center}
    \includegraphics[width=14cm]{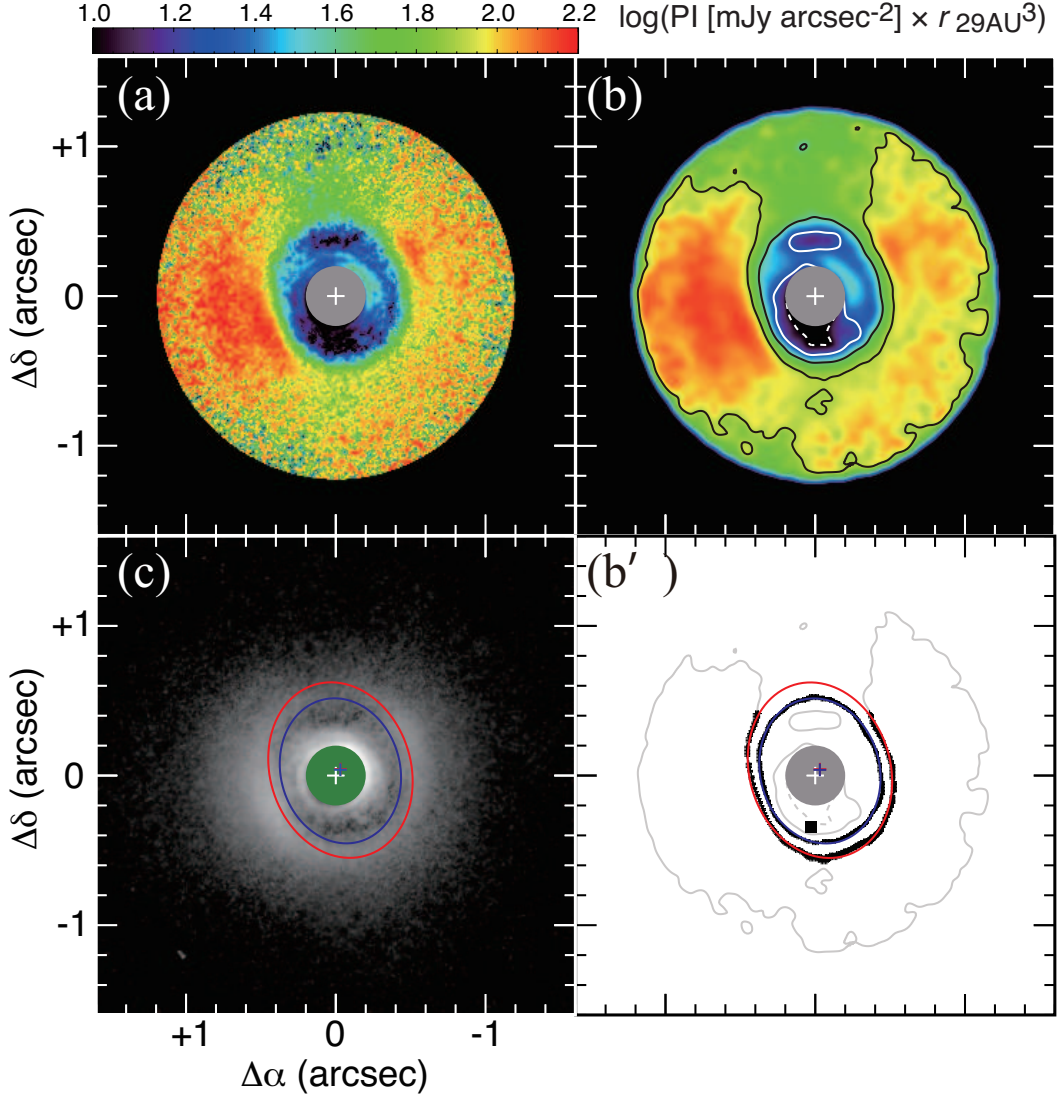}
  \end{center}
\caption{(a) The image in $r\leq 1\farcs1$ of ``modulated PI'', which is the PI 
multiplied by the dimensionless parameter $r_{\mathrm{29AU}}^{3}$, 
where $r_{\mathrm{29AU}}$ is the radius normalized by 29 AU. White cross and 
gray circle shows the stellar position and the regions affected by the mask, 
respectively. 
(b) The image of PI multiplied by  $r_{\mathrm{29AU}}^{3}$ after the convolution
with a Gaussian function whose FWHM is 8 pixels, $0\farcs072$.  
The contours of 10 and 20 mJy arcsec$^{-2}$ are shown in 
white broken lines and white solid lines, and the levels of 
40 and 80 mJy arcsec$^{-2}$ are shown in black solid lines. 
(b') The results of elliptical fittings to the contours of 40 mJy arcsec$^{-2}$ 
(\textit{blue}) and 
80 mJy arcsec$^{-2}$ except $-45^{\circ} \leq \mathrm{PA} \leq 40^{\circ}$ (\textit{red}). 
The centers of the ellipses are indicated by blue and red crosses, though the latter 
is hard to see because they almost perfectly agree with each other (see Table 1). 
The pixels used in the fitting are shown in black crosses. 
(c) The PI image, same as figure \ref{polimage}a, superposed on the results 
of the elliptical fittings. }
\label{pi-modbyr3}
\end{figure}

To clearly identify these asymmetric features in the image, 
figure \ref{pi-modbyr3} shows the map of ``modulated PI'', the PI 
normalized by $r_{\mathrm{29AU}}^{-3}$, 
where $r_{\mathrm{29AU}}$ is a dimensionless radius normalized by 29 AU.
Two distinct power-law regions, divided by TZ where the contours are
crowded, can be recognized. 
As indicated by the normalized intensities in 
equations (\ref{eqn:pl1}) and (\ref{eqn:pl2}), 
the average of modulated PI in the inner and outer power-law 
regions should be $\approx 21$ mJy arcsec$^{-2}$ and $\approx 96$ mJy 
arcsec$^{-2}$ (see also figure \ref{pi-rprofile}c). 
However, the variation of modulated PI in the inner 
power-law region is quite remarkable, ranging from 10 to 40 mJy arcsec$^{-2}$. 
This is significant compared to the uncertainties estimated from the 
original PI map (3$\sigma$ = 0.5 mJy arcsec$^{-2}$ in the original PI  
corresponds to $\approx 7.8$ mJy arcsec$^{-2}$ in modulated PI 
at $r=0\farcs5$). 
These asymmetries in the inner power-law region 
could be related to the perturbation by a protoplanet 
inside the inner ring \citep{biller14,reggiani14}.
Another intriguing feature is that the location of deep minimum 
in the southern part of the inner power-law region seems to coincide 
with the point source at $\lambda = 7$ mm revealed with J-VLA, 
as indicated in figure \ref{pi-modbyr3}b$^\prime$. 
\citet{osorio14} speculate that it
might originate from the circumplanetary disk associated 
with a protoplanet inside the gap, which is also identified 
in thermal emission at $\lambda =7$ mm. 
The relation between these features 
and the possible protoplanet in TZ will be 
discussed further in \S 4.2. 

To quantitatively describe the shape of TZ, 
elliptical fitting is made to the contours of 40 and 
80 mJy arcsec$^{-2}$ in modulated PI;
the results are presented in table \ref{tbl-1}.  These ellipses well delineate 
the edges of TZ in the original PI map, as shown 
in figure \ref{pi-modbyr3}c. The major axes of the ellipses nearly 
agree with the disk major axis (PA$=5^{\circ}$), 
and the ratio of minor to major axes is $\approx 0.8$, which is smaller than 
the case when an infinitesimally thin disk is 
observed from the inclination angle of $13^{\circ}$. 
These are consequences of the asymmetry that the PI near the 
minor axis of the disk is brighter than near the major axis 
(see also figures \ref{polimage} and \ref{pi-azprofile}).
The centers of the ellipses agree with each other but  
are shifted $\approx 0\farcs047$ northwest from the stellar position,
caused by the east-west asymmetry along the minor axis in 
the outer power-law region. 

\begin{table}
\caption{Results of elliptical fitting in figure \ref{pi-modbyr3}.}
\label{tbl-1}
\begin{center}
\begin{tabular}{cccccc}
      \hline
contour & major & minor & $\Delta \alpha$\footnotemark[a] 
& $\Delta\delta$\footnotemark[a] & PA \\ 
 (mJy~arcsec$^{-2}$) & axis ($\arcsec$) & axis ($\arcsec$) & ($\arcsec$) & ($\arcsec$) & ($^{\circ}$) \\
      \hline
40  & 0.490 & 0.399 & $-$0.030  & $+0.033$ & 14.2 \\
80\footnotemark[b] & 0.596 & 0.471 &  $-$0.031 
& $+0.038$ & 16.8 \\
\hline
\end{tabular}
\end{center}
\end{table}
\footnotetext[a]{Offset from the stellar position }
\footnotetext[b]{Only in $40^{\circ} \le \mathrm{PA} \le 315^{\circ}$. }

\section{Discussion}

In the following 
subsections, we discuss observed features of the disk from larger to smaller 
size scales in order; a double power-law distribution seen in the  
radial profile of PI is discussed in \S \ref{subsec-disc1}, 
possible presence of a protoplanet in the transition zone (TZ) is examined in 
\S \ref{subsec-disc2},  and asymmetries on smaller azimuthal scales are 
mentioned in \S \ref{subsec-disc3}.
It is impossible to derive the realistic 
internal structure of the disk solely from the PI image in $H$-band. 
Existence of gap between $40-70$ AU in radius has been inferred by 
recent J-VLA observations \citep{osorio14}, which provided us another 
important information about the disk internal structure. 
The emission at $\lambda=7$~mm, however, is mainly arising from 
dust grain with a larger size, and we cannot make any combined analysis of these 
two images unless we put further assumptions on dust size distribution 
as well as on size-dependence of dust spatial distribution. We therefore 
decide in this paper to interpret $H$-band image with a simple disk 
model whose surface density and temperature distributions ($\Sigma$ and $T$) 
have a power-law and the vertical density distribution is determined by 
hydrostatic equilibrium (Appendix \ref{appendix1}). 
Although our disk model is simple, it is useful for 
analyzing the location of scattering surface, the key to 
interpret the PI image in $H$-band. As explained in the 
Appendix \ref{appendix1}, $\Sigma$ essentially parameterizes the volume 
emissivity of disk material while $T$ parameterizes the 
vertical distribution of disk material. 
Further analysis on disk structure based on a more realistic model 
should be made after sensitive observations of dust emission 
at millimeter/sub-millimeter wavelengths are made with ALMA 
in angular resolution similar to this study.

\subsection{Disk structure that reproduces a double power-law 
distribution of PI's radial profile \label{subsec-disc1}}

An azimuthally-averaged radial profile of the PI in $H$-band 
shows a double power-law distribution, and the exponents in
the both power-law regions are $-3$ (figure \ref{pi-rprofile}).
The same exponent was also obtained in 
total intensity at $\lambda =1.1~\mu$m and $1.6~\mu$m 
\citep{grady07,fukagawa10}. 
A geometrically thin but optically thick disk model 
(e.g., \cite{whitney92})
has been frequently referred to as an 
analytical explanation for an $r^{-3}$-distribution 
(e.g., \cite{wis08}), 
but this is not the only explanation for such radial dependence.
\citet{grady07} showed that a model disk with a pressure scale height 
of $H(r) \propto r^{+1.065}$ can reproduce the $r^{-3}$-distribution 
of surface brightness at $\lambda =1.1 ~\mu$m by 
using 3D Monte Carlo radiative transfer code 
\citep{whitney03a,whitney03b,whitney04}. It was also reported that such a 
minimally flaring disk (i.e., the aspect ratio $H(r)/r$ is independent of 
$r$)\footnote{Such a case has been referred to as a ``flat disk'' 
in some studies (e.g., \cite{chiang97}).} 
with surface density $\Sigma(r) \propto r^{-1}$
can explain not only the radial distribution of scattered light 
but also the visibility amplitudes at $\lambda =1.3$ mm \citep{panic08} and
the SED at longer wavelengths \citep{meeus10, honda12}.

In fact, $r^{-3}$-dependence of the disk emission may be a natural 
consequence by the combination of $\Sigma(r) \propto r^{-1}$
and $H(r)\propto r$. 
The scattering surface of a disk corresponds to the positions 
where the optical depth measured radially from the star 
is $\approx 1$, while the energy flux of incident stellar 
radiation can be scaled by $\beta/r^{2}$, where $\beta$ is 
the grazing angle at the scattering surface. 
It can be derived analytically that 
$\beta$ is nearly proportional to $r^{-1}$ for an axisymmetric 
disk when $\Sigma(r) \propto r^{-1}$ and $H(r) \propto r$, 
accounting for the $r^{-3}$-brightness distribution (Appendix \ref{appendix2}). 
This means that 
the double power-law distribution 
(figures \ref{pi-rprofile} and \ref{pi-modbyr3})
is realized if $\beta$ 
is proportional to $r^{-1}$ in each power-law region 
but its value in the inner region is significantly smaller 
than that in the outer region. 
It can happen when the height of the scattering surface 
normalized by the radius in the inner region is smaller
than that in the outer region. 

\begin{figure}
  \begin{center}
    \includegraphics[width=12cm]{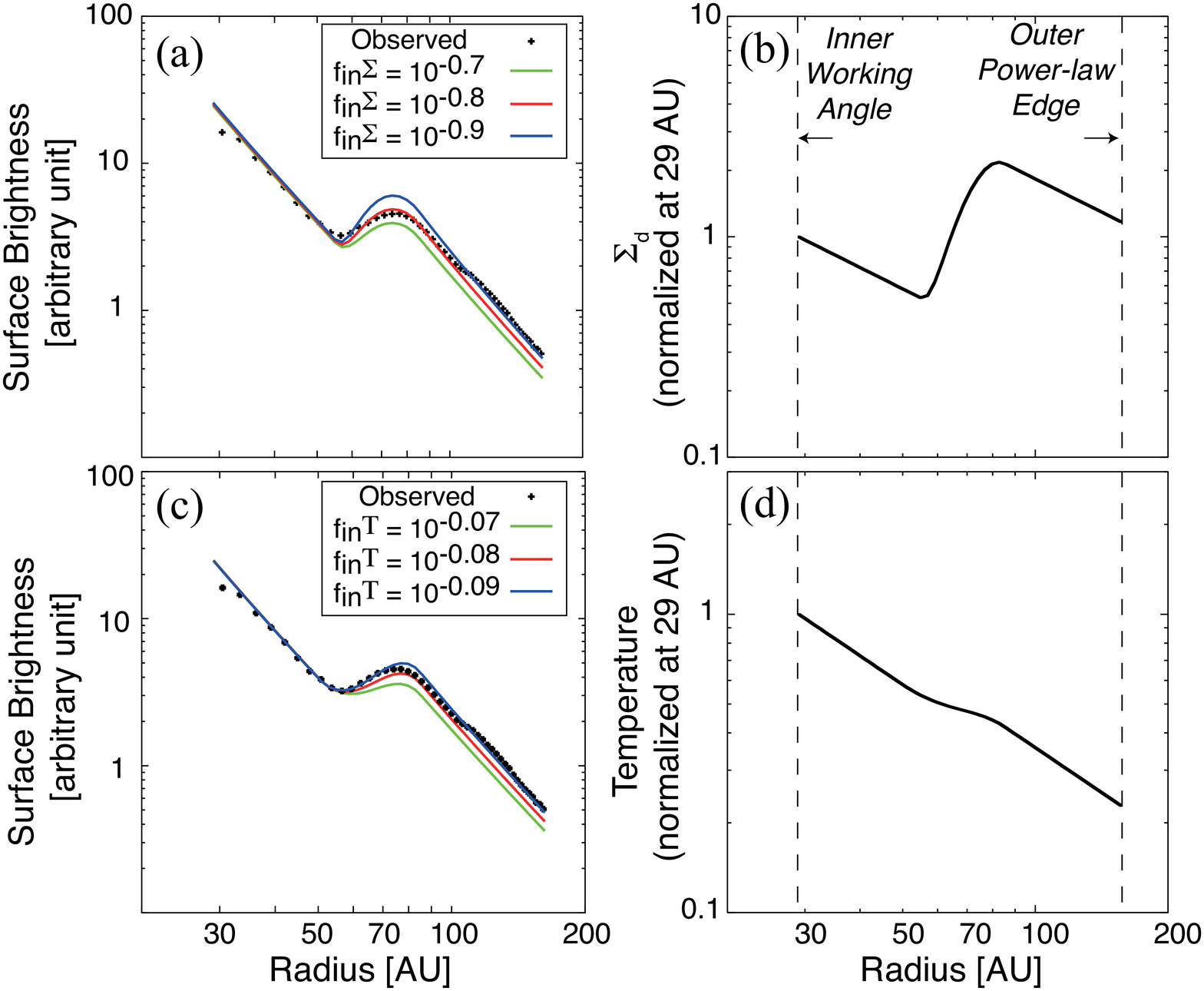}
  \end{center}
\caption{Models calculation that reproduce the 
averaged radial profile (\textit{black crosses}) shown in figure \ref{pi-rprofile}. 
Surface density distribution $\Sigma _{\mathrm{d}}(r)$ 
and scale height of the model disk $H(r)$ is proportional to $r^{-1}$ and $r^{+1}$, 
respectively, in $r=29-50$ AU and $r=85-160$ AU. 
$\Sigma_{\mathrm{out}}$ and $T_{\mathrm{out}}$ at 
$r_{\mathrm{out}}=85$ AU are set to be 9 g cm$^{-2}$ and 
53 K, but these choices are not important in the estimate for 
$f_{\mathrm{in}}^{~\Sigma}$ or $f_{\mathrm{in}}^{~T}$ (see Appendix for more detail). 
(a) The proportionality constant for $\Sigma _{\mathrm{d}}(r)$
is changed, as described in the equation (\ref{sdr}). The lines in 
green, red and blue represent the results of model calculations 
for $f_{\mathrm{in}}^{~\Sigma} = 10^{-0.7}, 10^{-0.8}$ and $10^{-0.9}$, 
respectively. (b) $\Sigma_{\mathrm{d}}(r)$ given by equations 
(\ref{sdr}) and (\ref{fdr}) when  $f_{\mathrm{in}}^{~\Sigma} = 10^{-0.8}$, 
normalized by the surface density at $r=29$ AU.
(c) The proportionality constant for $T(r)$
is changed, as described in the equations (\ref{sdr}) and (\ref{fdr}) 
but replace each $\Sigma$ with $T$. The lines in 
green, red and blue represent the results of model calculations 
for $f_{\mathrm{in}}^{~T} = 10^{-0.07}, 10^{-0.08}$ and $10^{-0.09}$, 
respectively. (d) $T(r)$  
when $f_{\mathrm{in}}^{~T} = 10^{-0.09}$, normalized by the temperature 
at $r=29$ AU.}
\label{pi-models}
\end{figure}

A simple example for such a case
is that the surface density distribution of dust particles responsible 
for the scattering in near infrared, $\Sigma_{\mathrm{d}}(r)$, is not  
a smooth power-law but has an irregularity. 
To roughly estimate the amount of irregularity in $\Sigma_{\mathrm{d}}(r)$, 
we compare the observed PI distribution with a simple model calculation 
whose details are described 
in Appendix \ref{appendix1}. 
As discussed later, \citet{wagner14} also carry out more 
elaborate modeling, but our simple model is helpful for 
analyzing the location of scattering surface. 
$\Sigma_{\mathrm{d}}(r)$ of the model disk is expressed by
\begin{equation}
\Sigma_{\mathrm{d}}(r) = f_{\mathrm{d}}(r)\Sigma_\mathrm{out} 
    \left(\frac{r}{r_\mathrm{out}}\right)^{-1},
     \label{sdr}
\end{equation}
where 
\begin{equation}
f_{\mathrm{d}}(r) = 
  \left\{
    \begin{array}{ll}
     f_{\mathrm{in}}^{~\Sigma} & (r\leq r_\mathrm{in}), \\
     f_{\mathrm{TZ}}(r)  & (r_\mathrm{in} < r< r_\mathrm{out}), \\
     1 & (r_\mathrm{out} \leq r).
    \end{array}
  \right.
\label{fdr}
\end{equation}
Here, $f_{\mathrm{in}}^{~\Sigma}~(0\le f_{\mathrm{in}}^{~\Sigma} \le 1)$ 
is a depletion factor 
for the surface density in the inner power-law region, 
$(r_\mathrm{in}, r_\mathrm{out})$ are the inner and outer boundaries of 
TZ and are set to be (50 AU, 85 AU) based on 
the fitting results expressed by equations (\ref{eqn:pl1}) and (\ref{eqn:pl2}),
and $\Sigma_\mathrm{out}$ is the surface density at $r_\mathrm{out}$.
The surface brightness distribution normalized by 
$r_{\mathrm{29AU}}^{-3}$ does not show any gap but smooth rise in TZ 
(figure \ref{pi-rprofile}c), and this cannot be reproduce by a model disk with
a dust devoid zone.  
To smoothly connect $\Sigma_{\mathrm{d}}(r)$ in TZ as well as at 
its boundaries, $f_{\mathrm{TZ}}(r)$ in equation (\ref{fdr}) is 
defined as a third order polynomial that simultaneously satisfies the 
following equations:
\begin{equation}
        \begin{array}{ll}
        f_{\mathrm{TZ}}(r_\mathrm{in}) &= f_{\mathrm{in}}^{~\Sigma}, \\
        f_{\mathrm{TZ}}(r_\mathrm{out}) &= 1, \\
        \frac{d}{dr} f_{\mathrm{TZ}}(r_\mathrm{in}) &= 0, \\
        \frac{d}{dr} f_{\mathrm{TZ}}(r_\mathrm{out})&= 0.
        \end{array}    
        \label{ftzr}
\end{equation}
We also assume that $T(r) \propto r^{-1}$ 
to set the pressure scale height $H(r) \propto r^{+1}$ 
(Appendix \ref{appendix1}; see also \cite{chiang97}), and the 
disk is isothermal and in hydrostatic equilibrium along the 
vertical axis. The multiple scattering effect is ignored in the
radiative transfer calculation for simplicity.
Figure \ref{pi-models}a shows the results of the model calculations. 
The $r^{-3}$-dependence is reproduced in 
both $r<50$ AU and $r>85$ AU regions, but 
a larger proportionality constant for $\Sigma_{\mathrm{d}}(r)$
in the outer region makes the location 
of the scattering surface higher from the mid-plane, resulting in larger 
$\beta$ and brighter PI. Judging from figure \ref{pi-models}a, 
$f_{\mathrm{in}}^{~\Sigma}= 10^{-0.8} (\approx 0.16)$ seems to
give the best fit to the observed contrast between the two 
power-law regions.  
$\Sigma_{\mathrm{d}}(r)$ given by equations (\ref{fdr}) and (\ref{ftzr}) 
with $f_{\mathrm{in}}^{~\Sigma} = 10^{-0.8}$ is presented in figure \ref{pi-models}b. 
There seems a gap in $r\approx 40-70$ AU 
with the minimum at $r\approx55$ AU as suggested by previous studies 
\citep{quanz13,osorio14}, though the profile shape is qualitatively different 
from a planetary gap having a characteristic size scale of the width 
(e.g., \cite{zhu11}). Despite many simplifications, the estimate 
by our model is roughly consistent with the results 
obtained by \citet{wagner14}; $f_{\mathrm{in}}^{~\Sigma} \approx 0.3$ 
in the model disk with $H(r) \propto r^{+1.3}$.

However, the above depletion factor for 
the inner power-law region should be regarded as the lower limit, 
i.e.,  $f_{\mathrm{in}}^{~\Sigma}$ should be $\ge 0.16$. 
This is because irregularity in radial temperature distribution 
can also produce the observed double power-law distribution of 
PI, even without any irregularity in $\Sigma_{\mathrm{d}}(r)$ 
like equations (\ref{fdr}) and (\ref{ftzr}); 
if the temperature proportionality 
constant is slightly higher in the outer power-law region, 
it makes the disk aspect ratio $h\equiv H(r)/r$ and 
$\beta$ in this region also larger. 
Our simple model similar to the above one but 
each $\Sigma$ in equations (\ref{sdr}) - (\ref{ftzr}) is
replaced by $T$ reproduces the observed PI profile when 
$f_{\mathrm{in}}^{~T} = 10^{-0.09}-10^{-0.08} (\approx 0.81-0.83)$, corresponding 
to the change in 
$h$ by a factor of $0.90-0.91$ 
(see Appendix \ref{appendix2} for further details). 
This is again consistent with the model calculations by \citet{wagner14} 
who show that a scale height change of 
0.86 in the inner power-law region can reproduce the 
observed PI image in $H$-band\footnote{Another important conclusion 
obtained by \citet{wagner14} is that the apparent gap in 
PI at $40\le r \le 70$ AU cannot be reproduced solely by 
changing the scale height of the inner wall at $r\approx 25$ AU.}. 
There is a degeneracy 
between $f_{\mathrm{in}}^{~\Sigma}$ and $f_{\mathrm{in}}^{~T}$ 
in models that reproduce the PI image, as 
pointed out by \citet{wagner14}. 
The radial intensity profile at $\lambda = 7$ mm is reproduced 
by a disk model in which there is no significant discrepancy in dust 
surface density between $r=30-40$ AU and $r> 70$ AU \citep{osorio14}, 
which might favor a larger $f_{\mathrm{in}}^{~\Sigma}$. 
Further imaging studies, especially at millimeter and submillimeter 
wavelengths, will still be necessary to clearly disentangle this 
degeneracy.

\subsection{Possible presence of a protoplanet in the transition zone (TZ) \label{subsec-disc2}}

The change of $\beta$ 
described in \S 4.1 may be caused by 
interaction between a protoplanet and the disk. 
\citet{jcondell12}, for example, examine the 
detail disk structure under the influence of a forming planet and 
simulate the images at various wavelengths. They obtain 
a radial distribution of near infrared scattered light 
qualitatively similar to our observed results.
Such a double power-law distribution is produced because the 
protoplanet creates a gap in the disk initially having flared surface, 
resulting in larger $\beta$ on the scattering surface at outer radii. 
Furthermore, cooling within the trough and heating 
on the far wall alter the vertical density profile of the gap region
in such a way that the difference of $\beta$ is enhanced even more. 
In their simulations, the brightness contrast between 
the inside and outside of the gap is much larger (a factor of 
$\sim 10^2$ in normalized intensity even in the 
lowest contrast case) than our observational results,  
suggesting that the mass of a protoplanet in TZ 
is so low that it only creates a shallower gap 
than the cases of \citet{jcondell12}; they deal with 
the cases of planet masses much higher than 
the critical mass for a gap opening proposed by 
\citet{crida06}. 

More recently, \citet{osorio14} obtain the image of dust thermal emission 
at $\lambda =7$ mm with J-VLA. They detect ring-like emission delineating
the bright rim at $r\approx0\farcs2$ revealed by the $H$-band PI image 
with VLT/NACO  \citep{quanz13}. The azimuthally-averaged 
radial intensity profile at $\lambda =7$ mm gives a 
hint that there is an annular gap in $r=40-70$ AU from the star that 
coincides with the gap seen in the $H$-band PI images. 
Based on these results, \citet{osorio14} carry out disk modeling in which there is a
dust devoid zone in $r=40-70$ AU, 
and they successfully reproduce the radial intensity profile at $\lambda = 7$ mm.
Meanwhile, the scattered light in $H$-band is detected  
even in the gap (figures \ref{polimage} \& \ref{pi-rprofile}), 
indicating that there is 
some amount of small dust particles in this zone. 
These two observational results are compatible if 
larger dust particles responsible for thermal 
emission at millimeter wavelengths are more deficient 
in the gap region. 
In fact, previous studies on the motion of dust particles in a gas disk 
under the influence of a planet show that particles with a modest
size (i.e., millimeter and centimeter in size) tend to be removed 
more quickly compared to the smaller particles during the 
formation of a planetary gap near the orbit \citep{paar06,lyra09,muto09}. 

The J-VLA image at $\lambda =7$ mm also reveals a faint point source, 
$0\farcs34$ at PA=$175^{\circ}$ from the central star. 
\citet{osorio14} discuss 
the possibility that it arises from thermal dust radiation of a circumplanetary 
disk associated with the protoplanet sculpting the gap. 
In the image of modulated PI in $H$-band, on the other hand, 
there is a deep minimum in the inner power-law region whose 
location seems to coincide with the point source at $\lambda = 7$ mm
(figure \ref{pi-modbyr3}b$^\prime$; see also figure \ref{pi-azprofile}a). 
This can be regarded 
as another sign of the protoplanet, because 
the disk scattering surface can be 
lowered by the protoplanet's gravity, 
making an observable dip in near infrared scattered light \citep{jcondell09}. 

To constrain the planet mass, 
it is important to precisely measure the depth and width of the 
gap and to compare them with a latest theoretical model of gap structure 
(e.g., \cite{duffell13,kanagawa14}). 
An analytic formula among the gap depth, planet mass, disk aspect 
ratio and viscosity is presented in a separate paper \citep{kanagawa15}, 
in which the formula is applied to the gap depth inferred from 
J-VLA observations of HD 169142 by \citet{osorio14}. 
Imaging studies of both dust and gas emission
with the ALMA will also be important in near future. 

\subsection{Asymmetric features in the outer power-law regions \label{subsec-disc3}}

As mentioned above, 
there are remarkable asymmetries both in the inner and outer 
power-law regions. 
Significant asymmetric variation in the inner power-law 
region (figures \ref{pi-azprofile} and \ref{pi-modbyr3}) 
could be due to the perturbation by 
a protoplanet inside the inner ring 
\citep{biller14,reggiani14}.  The deep minimum of PI in $H$-band in the 
inner power-law region coincides with the point source at $\lambda=7~$mm, 
which can be regarded as another sign of a protoplanet surrounding the 
circumplanetary disk. The azimuthal distributions at the outer radii 
tend to have the 
maxima at PA$\approx 110^{\circ}$, along the disk minor axis east to the star.

\begin{figure}
  \begin{center}
    \includegraphics[width=14cm]{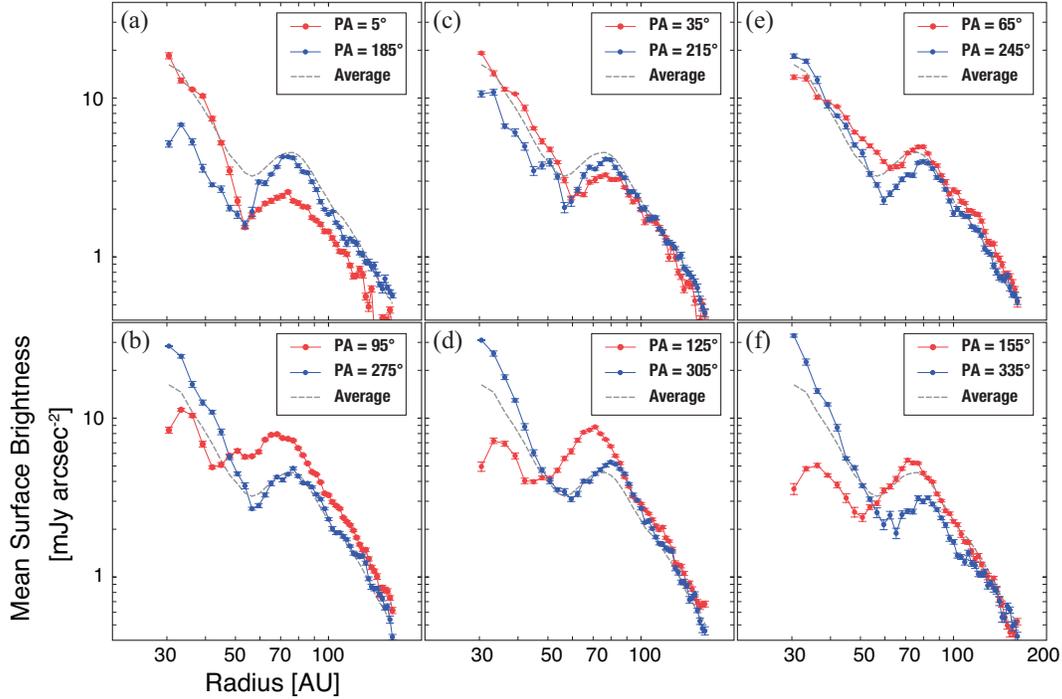}
  \end{center}
  \caption{Radial profiles of the PI along the pairs separated by $180^{\circ}$ in PA. 
(a) PA = $5^{\circ} \pm 5^{\circ}$ in red and PA = $185^{\circ} \pm 5^{\circ}$ in blue. Same as 
figure \ref{pi-majmin}a.
(b) PA = $95^{\circ} \pm 5^{\circ}$ in red and PA = $275^{\circ} \pm 5^{\circ}$ in blue. Same as 
figure \ref{pi-majmin}b.
(c) PA = $35^{\circ} \pm 5^{\circ}$ in red and PA = $215^{\circ} \pm 5^{\circ}$ in blue. 
(d) PA = $65^{\circ} \pm 5^{\circ}$ in red and PA = $245^{\circ} \pm 5^{\circ}$ in blue. 
(e) PA = $125^{\circ} \pm 5^{\circ}$ in red and PA = $305^{\circ} \pm 5^{\circ}$ in blue. 
(f) PA = $155^{\circ} \pm 5^{\circ}$ in red and PA = $335^{\circ} \pm 5^{\circ}$ in blue. 
Radial profile averaged over the whole direction (shown in figure \ref{pi-rprofile})
is indicated by gray broken line in each panel. }
\label{pi-pairs}
\end{figure}

Besides these features, there are small-scale asymmetries in the 
outer power-law region (see figure \ref{pi-modbyr3}), 
though the amount of variation 
is smaller than in the inner power-law region. 
There are two possibilities for the origin of these 
small-scale asymmetries. One is 
corrugation of the scattering surface 
in the outer region, causing a small-scale variation in grazing angle  
($\beta$). Such corrugation may be due to local fluctuations
in disk surface density that may affect the self-gravity of the disk 
and hydrostatic balance in its vertical direction \citep{muto11}.
The other possibility is a shadowing effect of 
the structure in the inner power-law regions. 
As described in \S 4.1, the radial intensity profile of near infrared 
scattered light as well as the broadband SED are basically 
explained by a minimally flaring disk 
(see also \cite{grady07,panic08,
meeus10,honda12,wagner14}). 
If a part of the inner structure of such a disk is slightly 
puffed up, it may cast a shadow on the outer region observable
in the near infrared image. Radial profiles 
along the major and minor axes (figure \ref{pi-majmin}) 
actually show that 
the surface brightness in the inner ($r\leq 50$ AU) and the outer ($r\geq 80$ AU) 
regions  are anti-correlated with each other. 
Similar tendency is also seen 
in all the pairs of radial profile $180^{\circ}$ apart in PA from 
each other (figure \ref{pi-pairs}). 
These two possibilities may be disentangled by 
search for time variation on a shorter timescale in scattered-light 
images; if the asymmetries in the outer disk are caused by
shadowing of the inner region, asymmetric patterns in the outer disk 
may vary on a short timescale 
that synchronizes with the change of disk structure in the inner 
power-law region (see also \cite{wagner14} for discussion on 
the relationship between the time variation seen in SED and 
structural change at much inner radii as small as 
$r \sim 0\farcs3$). 

The PI in the outer region tends to be higher 
along the minor axis than the major axis (figure \ref{pi-majmin}). This is 
completely opposite to other disks showing
brighter PI along the minor axis \citep{muto12, kusakabe12}, 
which is naturally explained by the highest polarization degree 
in $90^{\circ}$ scattering. The observed features of the disk around 
HD 169142 are rather difficult to interpret in terms of simple Rayleigh 
scattering by an axisymmetric disk in nearly face-on geometry. 
It is more likely that the most of these asymmetric features 
reflect real disk structure on smaller scales, as discussed above. 

\section{Conclusions}

We carry out coronagraphic observations of the circumstellar disk 
around HD 169142 in $H$-band polarized intensity (PI) 
with Subaru/HiCIAO. The emission scattered by dust 
particles at the disk surface in $0\farcs2 \leq r \leq 1\farcs2$,
corresponding to $29 \leq r \leq 174$ AU at the distance to 
HD 169142, is successfully detected. 
Our conclusions are summarized as follows: 

\begin{enumerate}

\item The overall structure of the PI is nearly axisymmetric. 
The azimuthally-averaged radial profile of the PI shows 
a double power-law distribution, in which the PIs in 
$r=29-52$ AU and $r=81.2-145$ AU respectively show 
$r^{-3}$-dependence. These two power-law regions are connected 
smoothly with a transition zone (TZ).  
The apparent gap in $r=40-70$ AU, previously reported 
by \citet{quanz13} and \citet{osorio14}, 
corresponds to the outer part of the inner power-law region and TZ. 

\item The PI image also contains asymmetric features. The variation 
along azimuthal direction in each annulus is systematically larger at 
inner radii, and the amount of variation in the inner power-law region 
is greater than 4. This could be due to the perturbation by a 
protoplanet inside the inner ring \citep{biller14,reggiani14}. 
The azimuthal distributions at the outer radii tend to have the 
maxima at PA$\approx 110^{\circ}$, implying that the eastern part is the near side 
of the disk and is brighter than the western part because of the 
forward scattering. 

\item The radial profile of $r^{-3}$-dependence is realized by the 
combination of surface density $\Sigma(r) \propto r^{-1}$
and the pressure scale height $H(r)\propto r$. 
The observed double power-law distribution
of the PI can be reproduced when the proportionality 
constant of surface density in the inner power-law region is smaller than that in 
the outer power-law region. 
Simple model calculation shows that the observed PI
profile matches the case of 
$f_{\mathrm{in}}^{~\Sigma}= 0.16$, where $f_{\mathrm{in}}^{~\Sigma}$
is the depletion factor for the proportionality 
constant of surface density in the inner power-law region. 
However, this should be regarded as a lower limit because irregularity in
temperature can also produce a double power-law distribution. 
Even with no irregularity in surface density, 
the observed PI profile is reproduced when 
$f_{\mathrm{in}}^{~T}=0.81-0.83$, where $f_{\mathrm{in}}^{~T}$ is the 
proportionality constant of 
temperature in the inner power-law region relative to the outer region. 
These results are consistent with more 
elaborate modeling by \citet{wagner14}. 
The degeneracy 
between $f_{\mathrm{in}}^{~\Sigma}$ and $f_{\mathrm{in}}^{~T}$ 
in models that reproduce the PI image should be disentangled by
further imaging studies, especially at millimeter and submillimeter 
wavelengths. 

\item The modeling based on the image at $\lambda =7$ mm 
\citep{osorio14} requires a dust devoid zone between $r=40-70$ AU, while 
the scattered light in $H$-band is significantly detected even in this gap. 
These observations suggest that larger dust particles responsible for thermal 
emission at millimeter wavelengths may be more deficient 
in the gap, as predicted by previous theoretical studies. 

\item The PI in the inner power-law region shows a deep 
minimum whose location seems to coincide with the point 
source at $\lambda = 7$ mm. 
This can be regarded as another sign of a protoplanet, because 
the scattering surface of the disk can be 
lowered by the protoplanet's gravity, making an observable dip in near 
infrared scattered light \citep{jcondell09}. 

\item The obtained PI image also shows small scale asymmetries 
in the outer power-law region. 
Possible origins for these asymmetries include corrugation of the
scattering surface in the outer region, and shadowing effect by a puffed
up structure in the inner power-law region. 
These two may be disentangled by 
search for time variation on a shorter timescale in the 
scattered-light image.

\end{enumerate}

\bigskip

We are grateful to the referee for valuable 
comments that improved our manuscript. 
This work is supported by 
MEXT KAKENHI Nos. 23103004 and 23103005,  
and by the U.S. National Science Foundation 
under Award No. 1009203.
This research has made use of the SIMBAD database,
operated at CDS, Strasbourg, France. 
Part of this research was carried out during the 
workshop ``Recent Development in Studies of 
Protoplanetary Disks with ALMA'', hosted by 
Institute of Low Temperature Science, 
Hokkaido University.

\appendix
\section{Description of the model calculations \label{appendix1}}

The model is based on a disk   
where the surface density and temperature 
distributions are given by power-law forms: 
\begin{equation}
 \Sigma = \Sigma_{\mathrm{out}} \left( \frac{r}{r_{\mathrm{out}}} \right)^{-p},
 \label{app:sigma}
\end{equation}
\begin{equation}
 T = T_{\mathrm{out}} \left( \frac{r}{r_{\mathrm{out}}}\right)^{-q},
  \label{app:t}
\end{equation}
where $\Sigma_{\mathrm{out}}$ and $T_{\mathrm{out}}$ are the surface density and 
temperature at $r_{\mathrm{out}}$. 
In the model, 
the sound speed $c_{\rm s}$ is approximated by
\begin{equation}
 c_{\rm s} = 10^{5} \left( \frac{T}{300\mathrm{K}} \right)^{1/2} 
  \mathrm{cm~s}^{-2}
    \label{app:cs}
\end{equation}
for simplicity. 

The disk is assumed to be vertically isothermal and in hydrostatic
balance with the gravity of the central star.  The density profile is
then expressed by 
\begin{equation}
 \rho(r,z) = \rho_{\rm c}(r) \exp 
  \left( -\frac{z^2}{2H^2} \right),
  \label{dens_rz}
\end{equation}
where $\rho_{\rm c} (r)$ is the midplane density and 
$H=c_{\rm s}/\Omega$  is the pressure scale height. 
 Assuming the Keplerian rotation, it is easy to see that 
$H \propto r^{(3-q)/2}$ is realized from equations (\ref{app:t}) and 
 (\ref{app:cs}). 
Using equation (\ref{app:cs}), 
the disk aspect ratio $h=H/r$ scales with 
\begin{equation}
 h = 0.1 \left( \frac{r}{100\mathrm{AU}} \right)^{(1-q)/2}\left(\frac{M_*}{M_{\odot}}\right)^{-1/2}.
 \label{app:h}
\end{equation}
The midplane density, on the other hand, is related with the surface density as 
\begin{equation}
 \rho_{\rm c} = \frac{\Sigma}{\sqrt{2\pi}H} \propto r^{-(2p-q+3)/2}.
 \label{app:rho0}
\end{equation}
Since our goal in this study is to assess the location of 
the scattering surface in a simple setup, we do not care about the 
consistency between the temperature and density
profile in terms of the energy balance. 

In radiative transfer calculation, 
we assume a single and 
isotropic scattering.\footnote{Thermal emission is included in the code as well,
but it is not significant as long as we consider the intensity
in near infrared beyond 10~AU.}  
The equation to be solved is
\begin{equation}
 \frac{dI}{dZ} = -\sigma W(r,z) B(T_{\ast}) 
  \exp\left[ -\tau_{\rm rad}(Z) - \tau(Z) \right],
  \label{transfer_scat}
\end{equation} 
where $\sigma$ is the scattering coefficient, 
$Z$ is the coordinate along the line of sight, 
$W(r,z)=1/4(r^2+z^2)$ is the geometric dilution factor, 
$\tau_{\rm rad}$ is the optical depth between the central star and the
location of the disk, and $\tau(Z)$ is the optical depth along the line
of sight.  In integrating equation (\ref{transfer_scat}), we first
calculate $\tau_{\mathrm{rad}}$ at all the location of the disk in the
grid and then integrate over line of sight from the observer position to
the disk. 

To compare with the observed radial profile of the PI, 
we modify the surface density or temperature distribution as 
described in \S 4. The averaged PI is assumed to be 
proportional to the averaged total intensity, which seems 
reasonable because the total intensity 
has a similar radial profile to PI 
\citep{grady07,fukagawa10}. We treat the unit of calculated intensities 
as an arbitrary one partly because of the uncertainty 
in polarization efficiency, and 
we fit the shape of the PI radial profile to estimate 
$f_{\mathrm{in}}^{~\Sigma}$ or $f_{\mathrm{in}}^{T}$. 

In this paper, we only deal with the cases of $p=1$ and $q=1$ and 
look at how the observed radial profile can be reproduced. 
The $q=1$ case corresponds to the disk whose aspect ratio $h$ 
is independent of $r$ as shown in equation (\ref{app:h}) \citep{grady07,panic08} 
and successfully reproduces the $r^{-3}$-dependence of the PI profile. 
The $r^{-3}$-dependence can also be derived analytically in 
$p=1$ and $q=1$ cases, as shown in Appendix \ref{appendix2}. 
Figure \ref{pi-models} shows the cases of $\Sigma_{\mathrm{out}} = 9$ g cm$^{-2}$, 
$T_{\mathrm{out}}= 53$ K at $r_{\mathrm{out}} = 85$ AU 
taken from equilibrium temperature 
for blackbodies directly exposed to the stellar radiation with 9.4 $L_{\odot}$ \citep{meeus12}.
Although the fiducial grain temperatures are set to the local blackbody
temperature, the grains are assumed to be highly reflective, with 
the scattering albedo of $\omega=0.9$, and 
their total extinction coefficient is set to be $\chi = 100$ cm$^{2}$ g$^{-1}$ 
($\sigma = \chi \omega$). 
The change of $\Sigma_{\mathrm{out}}$ or $T_{\mathrm{out}}$, however, 
almost scales the overall profile unless the disk gets optically thin, 
and it does not significantly affect the estimate for 
$f_{\mathrm{in}}^{~\Sigma}$ or $f_{\mathrm{in}}^{~T}$. 

The position of the scattering surface depends on two factors: 
(i) the volume emissivity of disk material determined by its density and emissivity, 
and (ii) the vertical distribution of disk material. The model described here 
essentially parameterizes (i) with $\Sigma$ and (ii) with $T$, respectively. 
The change of $\Sigma$ can be interpreted not only by the 
change of material density but also by the spatial variation of optical 
properties of dust particles. The change of $T$, on the other hand, can be 
related not only to the temperature, or the pressure scale height $H=c_s/\Omega$, 
but also to the degree of dust settlement toward the disk mid-plane. 
Since it is impossible to disentangle all these factors solely from the PI 
image in $H$-band, we do not examine the detailed disk structure based 
on a realistic disk model in this study. Further detailed analysis on disk 
structure should be made after sensitive observations of dust emission 
at millimeter/sub-millimeter wavelengths are carried out 
with ALMA in a similar angular resolution. 

\section{Analytic consideration for the cases of $p=1$ and $q=1$\label{appendix2}}

The results of the model calculation show that the cases of 
$p=1$ and $q=1$ successfully reproduce the $r^{-3}$-dependence of 
surface brightness. 
In fact, it is possible to predict
the radial profile of the surface brightness analytically in these 
cases as follows. 

The approximate solution to equation (\ref{transfer_scat})
is given by
\begin{equation}
 I \propto \beta/r^2,
 \label{scat_approx} 
\end{equation}
where $\beta$ is the grazing angle of the stellar light at the disk
surface.  Here, the disk scattering surface is determined by 
the height $z_{\rm s}$ at
which the optical depth from the central star becomes unity.
The optical depth $\tau_{\rm rad}$ can be calculated by
\begin{equation}
 \tau_{\rm rad} (r,z) = \int_{r_{\rm in}}^{r} 
  \chi \rho(r^{\prime},z) \frac{dr^{\prime}}{\cos{\theta}},
\end{equation}
where $\theta$ is given by
\begin{equation}
 \tan \theta = \frac{z}{r}.
\end{equation}
Substituting the density profile given by equation (\ref{dens_rz}), we
have 
\begin{equation}
  \int_{r_{\rm in}}^{r} 
  \chi \rho(r^{\prime},z) \frac{dr^{\prime}}{\cos{\theta}}
  = \int_{r_{\rm in}}^{r}
  \chi \rho_{\rm c}(r^{\prime}) 
  \exp \left[ -\frac{\tan^2 \theta}{2 h(r^{\prime})^2} \right] 
  \frac{dr^{\prime}}{\cos \theta}.
\end{equation}
Here, we have assumed that the central star is at the origin and used
the fact that $z^{\prime} = r^{\prime} \tan \theta$ along the line
connecting between the location of the disk considered and the stellar
position.  If we assume the disk with the constant $h$, the exponential
term is constant in the integration and therefore
\begin{equation}
 \tau_{\rm rad} (r,z) 
  = \chi \exp \left[ -\frac{1}{2h^2} 
	       \left( \frac{z}{r}\right)^2 \right] 
  \sqrt{1+\left(\frac{z}{r}\right)^2}
  \int_{r_{\rm in}}^{r} \rho_{\rm c} (r^{\prime}) dr^{\prime}.
  \label{opt_depth_gen}
\end{equation}
Finally, if the midplane density profile is given by
\begin{equation}
  \rho_{\rm c} (r) = \rho_{\rm c0} \left( \frac{r}{r_{\rm in}} \right)^{-2},
\end{equation}
its integration can be calculated as
\begin{equation}
  \int_{r_{\rm in}}^{r} \rho_{\rm c} (r^{\prime}) dr^{\prime}
   = \rho_{\rm c0} r_{\rm in}
   \left[ 1 - \left( \frac{r}{r_{\rm in}} \right)^{-1} 
   \right].
\end{equation}
Note that $\tau_{\rm rad}(r,z)$ has the form
\begin{equation}
 \tau_{\rm rad} (r,z) = f(r)g(z/r)\equiv f(r)g(u),
\end{equation}
where
\begin{equation}
 f(r)   = \rho_{\rm c0} r_{\rm in} 
   \left[ 1 - \left( \frac{r}{r_{\rm in}} \right)^{-1} 
   \right]
\end{equation}
and
\begin{equation}
 g(u) = \chi \exp \left[ -\frac{u^2}{2h^2} \right] 
  \sqrt{1+u^2}.
\end{equation}

We can define the disk scattering surface $z_{\rm s}(r)$ by 
\begin{equation}
 \tau_{\rm rad} (r,z_{\rm s}) = \tau_0,
\end{equation}
where $\tau_0$ is the value of the order unity.  
Then, the location of the scattering surface is given by
\begin{equation}
 \frac{z_{\rm s}}{r} \sim h \sqrt{\log C^2 
  - 2 \left( \frac{r_{\rm in}}{r} \right)}
  \label{app:zsr}
\end{equation}
where we have used $h \ll 1$ and
\begin{equation}
 C = \frac{\chi \rho_{\rm c0} r_{\rm in}}{\tau_0}.
\end{equation}
In the regions of $r \gg r_{\rm{in}}$, 
$z_{\rm s}/r$ in equation (\ref{app:zsr}) only
weakly depends on the radius, and the scattering surface always resides a few
times higher than the disk scale height.
Unlike a flared disk (e.g., \cite{chiang97}), the aspect ratio of the 
scattering surface, $z_s/r$, is almost constant 
(i.e., the scattering surface is nearly parallel to the radial direction from the star; 
see equation (\ref{app:zsr}) in Appendix \ref{appendix2}) when $p=1$ and $q=1$. 
In such a case, the surface brightness of the resultant scattering light is changed 
rather sensitively by the change of $\Sigma$. 

The grazing angle can be calculated by
\begin{equation}
 \beta = \frac{dz_{\rm s}}{dr} - \frac{z_{\rm s}}{r}.
\end{equation}
Since
\begin{equation}
 \frac{dz_{\rm s}}{dr} 
  = - \frac{\partial \tau_{\rm rad}/\partial r}{\partial \tau_{\rm
  rad}/\partial z}, 
\end{equation}
we have
\begin{equation}
 \beta = - \frac{r f^{\prime}(r) g(z_{\rm s}/r)}{f(r) g^{\prime}(z_{\rm s}/r)}
\end{equation}
where $^{\prime}$ denotes the derivative with respect to the argument.
Substituting the expressions of $f$ and $g$ functions and use $h \ll 1$, 
we obtain
\begin{equation}
 \beta \sim  \frac{h^2r_{\rm in}}{r}
  \left[\frac{z_{\rm s}}{r} \left(1-\frac{r_{\rm in}}{r}\right)\right]^{-1}.
  \label{beta-app2}
\end{equation}
In $r \gg r_{\rm in}$, the
last part in a bracket  is a weak function of $r$.
We therefore conclude that the grazing angle scales with
\begin{equation}
 \beta \propto r^{-1},
\end{equation}
and combining this with equation (\ref{scat_approx}), 
the radial profile of the scattered light is given by
\begin{equation}
 I \propto r^{-3}.
  \label{scat_prof_predict}
\end{equation}

Equation (\ref{beta-app2}) indicates that 
$\beta$ is nearly proportional to $h$ when $\Sigma$ distributes smoothly, 
while $h$ is proportional to $c_{\mathrm{s}}$, or $T^{1/2}$. 
The difference in observed proportionality constant of PI($r$) between 
$r < 50~\mathrm{AU}$ and $r > 85~\mathrm{AU}$ is $0.22$, 
as shown in equations (\ref{eqn:pl1}) and (\ref{eqn:pl2}). 
This difference corresponded to 0.47 in temperature if the aspect ratio $h$ remained  
constant at all radii. In our model, however, $h$ in $r < 50~\mathrm{AU}$ is 
smaller than that in $r > 85~\mathrm{AU}$, and the stellar radiation incident 
at the scattering surface at $r>85~\mathrm{AU}$ is brighter than the case of 
$h=$ constant at all radii. 
This is why the models with $f_{\mathrm{in}}^{~T}$ closer to the unity 
($=0.81-0.83$) in \S \ref{subsec-disc1} reproduces the observed radial profile. 


\end{document}